\documentclass[reprint,aps,pre,twocolumn,superscriptaddress,floatfix,showpacs,amsmath,amssymb]{revtex4-1}
\usepackage{verbatim}
\usepackage{graphicx,color,bbm, import}
\usepackage{epsfig}
\usepackage{amsmath}
\usepackage{bbold}
\usepackage{capt-of}
\usepackage{braket}
\usepackage{mathrsfs}  
\newcommand{\ketbra}[2]{\vert {#1} \rangle \langle{#2}\vert}
\newcommand{\grayeq}[1]{\textcolor[rgb]{0.7,0.7,0.7}{#1}}

\newcommand{\tcred}[1]{\textcolor[rgb]{0.7,0,0}{#1}}

\usepackage[normalem]{ulem}
\begin{document}
\title{Advances in quantum control of three-level superconducting circuit architectures}
\author{G. Falci}
\affiliation{Dipartimento di Fisica e Astronomia,
Universit\`a di Catania, Via Santa Sofia 64, 95123 Catania, Italy.}
\affiliation{CNR-IMM  UOS Universit\`a (MATIS), 
Consiglio Nazionale delle Ricerche, Via Santa Sofia 64, 95123 Catania, Italy.}
\affiliation{Istituto Nazionale di Fisica Nucleare, Via Santa Sofia 64, 95123 Catania, Italy.}
\author{P. G. Di Stefano}
\affiliation{Centre for Theoretical Atomic, Molecular and Optical Physics,
School of Mathematics and Physics, Queenâs University Belfast, Belfast BT7 1NN, United Kingdom.}
\affiliation{Dipartimento di Fisica e Astronomia,
Universit\`a di Catania, Via Santa Sofia 64, 95123 Catania, Italy.}
\author{A. Ridolfo}
\affiliation{Dipartimento di Fisica e Astronomia,
Universit\`a di Catania, Via Santa Sofia 64, 95123 Catania, Italy.}
\author{A. D'Arrigo}
\affiliation{Dipartimento di Fisica e Astronomia,
Universit\`a di Catania, Via Santa Sofia 64, 95123 Catania, Italy.}
\author{G. S. Paraoanu}
\affiliation{Low Temperature Laboratory, Department of Applied Physics,
Aalto University School of Science, P.O. Box 15100, FI-00076 Aalto, Finland.}
\author{E. Paladino}
\affiliation{Dipartimento di Fisica e Astronomia,
Universit\`a di Catania, Via Santa Sofia 64, 95123 Catania, Italy.}
\affiliation{CNR-IMM  UOS Universit\`a (MATIS), 
Consiglio Nazionale delle Ricerche, Via Santa Sofia 64, 95123 Catania, Italy.}

\begin{abstract}
Advanced control in Lambda ($\Lambda$) scheme of a solid state architecture of artificial atoms and quantized modes would allow the translation to the solid-state realm of a whole class of phenomena 
from quantum optics, thus exploiting new physics emerging in larger integrated quantum networks and for stronger couplings. 
However control solid-state devices has constraints coming from selection rules, due to 
symmetries which on the other hand yield protection from decoherence, 
and from design issues, for instance that coupling to microwave cavities is not 
directly switchable.  
We present two new schemes for the $\Lambda$-STIRAP control problem with the constraint of one or two  classical driving fields being always-on. We show how these protocols are 
converted to apply to circuit-QED architectures. We finally illustrate an application 
to coherent spectroscopy of the so called ultrastrong atom-cavity coupling regime.
\end{abstract}
\maketitle

\section{Introduction}
\label{sec:intro}
In the past decade several experiments have demonstrated 
evidence of multilevel coherence in superconducting 
artificial atoms (AAs)~\cite{ka:204-muraliorlando-prl-eit,ka:206-duttonorlando-prb-eit,ka:210-kellypappas-prl-cpt,ka:209-sillanpasimmonds-prl-autlertownes,ka:212-lihakonen-srep-qswitch,ka:216-xuhanzhao-natcomm-ladderstirap,ka:216-kumarparaoanu-natcomm-stirap}. 
Further exploiting coherence 
in such systems, and taking 
advantage of the inherent nonlinearity of multilevel
``atoms'' would allow 
important applications in solid-state quantum 
technologies~\cite{kb:210-nielsenchuang}, enabling tasks like multiqubit or multistate device processing~\cite{ka:211-timoney-nature-dressed,kr:211-younori-nature-multilevel,ka:209-lanyonwhite-natphys-quditprocessing} by adiabatic protocols, topologically protected computation~\cite{kb:209-pachos-topqcomp} or communication in distributed quantum networks\cite{ka:208-kimble-nature-qinternet,ka:214-darrigo-annals-hiddenent,ka:215-orieux-scirep-entrecovery,ka:207-darrigo-njp-memorychannel}. 
These are currently investigated roadmaps towards the design of fault tolerant complex quantum architectures~\cite{kr:210-ladd-nature-revqcomp,kr:213-devoretschoelkopf-science}, AAs being very promising since, compared to their natural counterparts, they allow for a larger degree of integration~\cite{kr:208-schoelkopf-nature-wiring,ka:214-machaustinov-ncomms-supmetamaterials,ka:214-mohebbicory-japph-arraymicrostrip,ka:216-brechtschoelkopf-natqinfo}, on-chip tunability, stronger couplings~\cite{ka:210-niemczyck-natphys-ultrastrong} 
and easier production and detection of signals in the novel regime of microwave quantum photonics~\cite{ka:213-nakamurayam-ieee-microwphot}.  

The major drawback of AAs, namely 
decoherence due to strong coupling to the electromagnetic~\cite{ka:191-falci-epl-elctrenvir,kr:201-makhlinschoenshnir-rmp} or the solid-state environment~\cite{ka:205-falci-prl-initial,ka:203-paladino-advssp-decoherence,kr:214-paladino-rmp},
has softened over the years~\cite{kr:213-devoretschoelkopf-science} yielding last-generation superconducting 
devices with decoherence times in the range 
$\sim 1-100\,\mu\mathrm{s}$~\cite{ka:211-bylander-natphys,ka:212-rigettisteffen-prb-trasmonshapphire,ka:214-sternsaclay-prl-fluxqubit3D}. Low decoherence, which however is achieved
at the expenses of reducing control resources, since protection of noise requires enforcing symmetries 
which imply selection rules. This for instance prevents to operate a three-level AA in the 
Lambda ($\Lambda$) scheme~\cite{kr:201-vitanov-advatmolopt} at the optimal working point where the qubit decoherence is minimal~\cite{ka:213-falci-prb-stirapcpb}. 
Recently it has been proposed 
an advanced control scheme which, combining the use of a two-photon pump pulse with a suitable slow modulation of the phase of the external fields, allows to operate dynamical $\Lambda$-scheme
protocols like $\Lambda$-STIRAP in last-generation AAs~\cite{ka:216-distefano-pra-twoplusone},
solving a problem raised in the 
last decade by several theoretical proposals~\cite{ka:205-liunori-prl-adiabaticpassage,ku:205-mariantoni-arXiv-microwfock,ka:206-siebrafalci-optcomm-stirap,ka:208-weinori-prl-stirapqcomp,ka:209-siebrafalci-prb,kr:211-younori-nature-multilevel}, which still awaits experimental demonstration. 

Our aim is to address the implementation in AAs with reduced control resources
of control schemes like $\Lambda$-STIRAP. This protocol is appealing because it implements with remarkable robustness a photon absorption-emission cycle which is a fundamental building block for processing in quantum architectures encompassing microwave resonating modes~\cite{ka:204-wallraff-superqubit,ka:203-plastinafalci-prb-communicating,ka:209-siebrafalci-prb,ka:213-muckerempe-pra-stirapsinglephgen}. In solid-state integrated systems a further limitation of control  emerges, since hardware fabrication does not allow to produce efficiently modulable couplings between 
individual devices. Switchable AA-cavity coupling can be achieved via control of the effective dynamics, either by standard modulation of detunings or by triggering multiphoton processes~\cite{ka:215-zeytinogluwallraff-phasetunable}. The former method allows to implement $\Lambda$-STIRAP with 
a single always-on classical field~\cite{ka:215-distefano-prb-cstirap}. 

In this work, after this introductory section where we introduce conventional $\Lambda$-STIRAP in three-level AAs, we present in \S\ref{sec:always-on} 
new schemes for the $\Lambda$-STIRAP control problem with the constraint of one or two  classical driving fields being always-on. In \S\ref{sec:circuitQED} we show how these protocols are 
converted to apply to circuit-QED architectures, yielding a key ingredient for several tasks 
with individual microwave photons, in systems where cavities cannot be coupled to AAs  
with switchable hardware. In \S\ref{sec:USC} we illustrate an application 
to coherent spectroscopy of the so called ultrastrong AA-cavity coupling regime~\cite{ka:210-niemczyck-natphys-ultrastrong,ka:212-scalari-science-USCTHz}. 
A discussion of results and perspectives are addressed in \S\ref{sec:conclusions}.

\subsection{Coherent population transfer in three-level atoms}
\label{sec:STIRAP}
STIRAP is an advanced control technique 
of $M>2$-level systems, 
that allows to transfer adiabatically population 
between two of the $M$ states, say $\ket{0}$ and $\ket{1}$, 
even in absence of direct coupling between them.
The presence of one or more intermediate states is necessary,  
but they are {\em never} populated. In this respect there 
are analogies with the Raman coupling, since both rely on  
two-photon processes. There is however a fundamental 
difference, namely while in the latter case 
small population of the virtual states requires 
far-detuned lasers, STIRAP relies  
on resonant external fields inducing destructive interference  
and coherent population transfer~\cite{ka:209-cohentannoudji-kosmos} is achieved adiabatically. 
This is a strong asset of  STIRAP, making it extremely remarkably robust and effective 
in producing faithful and selective population
transfer~\cite{kr:198-bergmann-rmp-stirap,kr:201-vitanov-annurev}.
\begin{figure}[t]\centering
\raisebox{20pt}{\includegraphics[width=0.14\columnwidth]{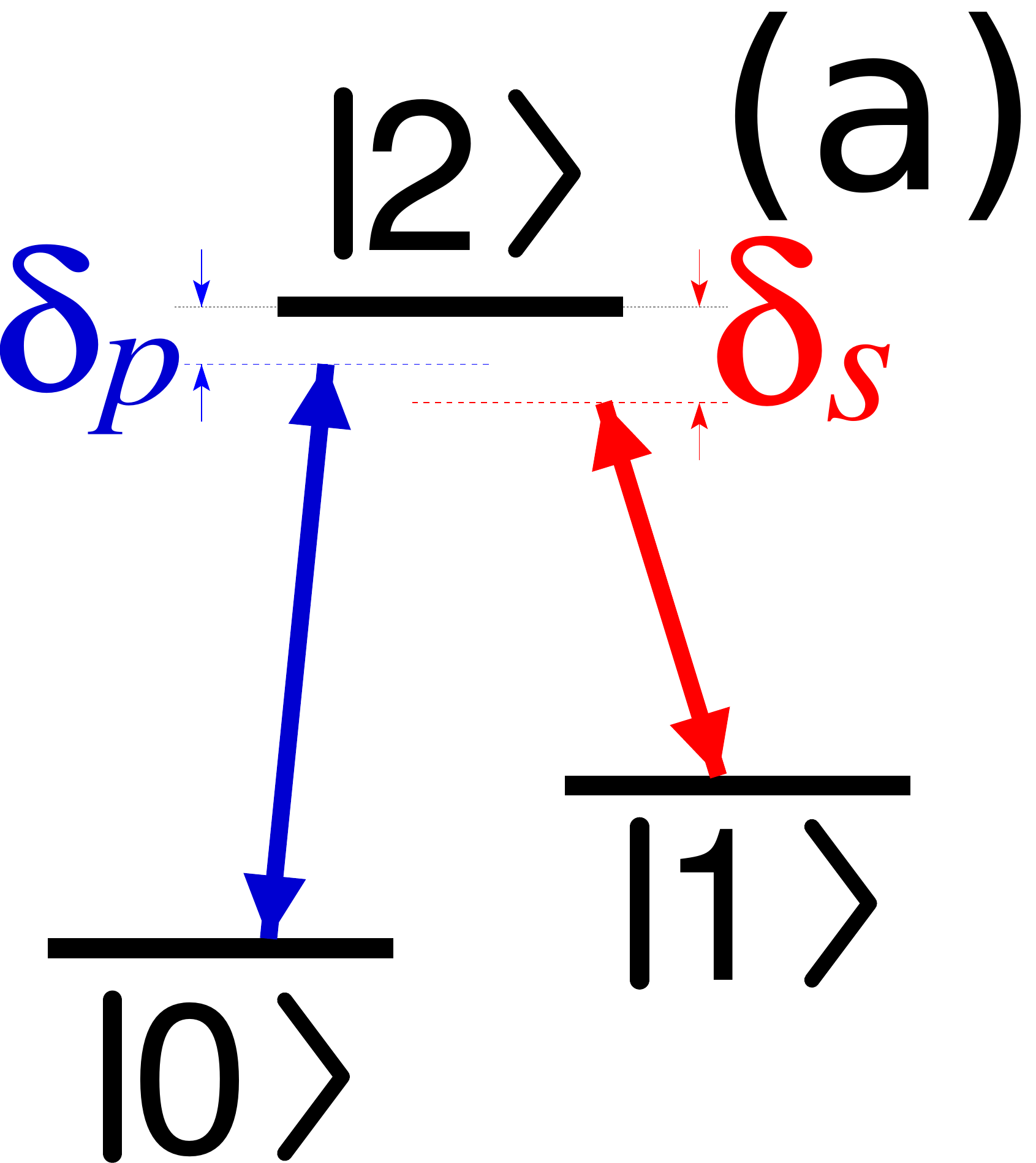}}
\includegraphics[width=0.42\columnwidth]{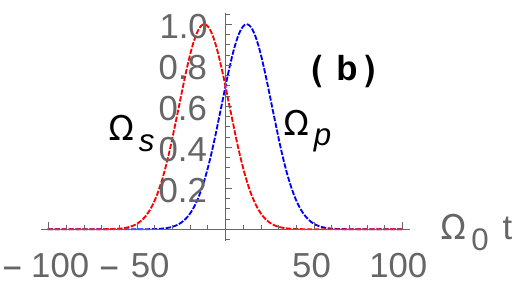}
\includegraphics[width=0.42\columnwidth]{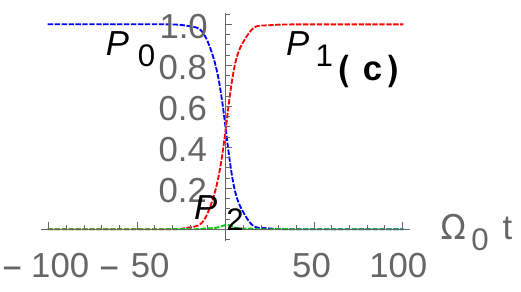}
\caption{(color online) (a) Three-level system driven with AC 
fields in $\Lambda$ configuration.
(b) External fields in the counterintuitive sequence: 
the Stokes field is switched on {\em before} the pump field 
(here $\Omega_0 T = 20, \tau= 0.6 \,T$).
(c) Population histories 
$P_{i}(t) = |\langle i |\psi(t) \rangle|^2$
for ideal STIRAP ($\delta = 0$): if 
the the Dark state evolves adiabatically, complete
population transfer $|0 \rangle \to |1 \rangle$ 
is achieved, while 
$|2 \rangle$ is never occupied.
\label{fig:stirap-ideal}}
\end{figure}

We consider a three-level AA with Hamiltonian
\begin{equation}
\label{eq:driven-H}
H:=H_0 + H_C(t)  
\end{equation}
where $H_0 := \sum_j \epsilon_j \ketbra{j}{j}$ models the 
undriven AA. The control 
$H_C = Q  \,A(t)$ is operated by a two-tone field 
$A(t)= \sum_{m=p,s}{\cal A}_{m}(t)  
\cos (\omega_m t)$, coupled to the operator ${Q}$, 
which is the analogous of the electric dipole for natural atoms.
The indirect linkage between ``bare'' 
levels $\{\ket{0},\ket{1}\}$ is provided by two external fields. The 
pump field has angular frequency $\omega_p \approx |\epsilon_2-\epsilon_0|$, 
triggers transitions $\ket{0} \leftrightarrow \ket{2}$ 
whereas the Stokes, $\omega_s \approx |\epsilon_2-\epsilon_1|$, triggers 
$\ket{0} \leftrightarrow \ket{2}$, yielding 
the $\Lambda$ configurations of  
Fig.~\ref{fig:stirap-ideal}a. 
Performing the Rotating Wave Approximation
(RWA) and expressing the approximate $H$ 
in a doubly rotating frame about the "bare" basis, we find 
\begin{equation}
\label{eq:H}
H = \begin{bmatrix}
0 								& 	0 							& \frac{1}{2}\Omega^\ast_p(t)	\\
0 								& 	\delta (t)					& \frac{1}{2}\Omega^\ast_s(t) 	\\
\frac{1}{2}\Omega_p(t) 		& 	\frac{1}{2}\Omega_s(t) 	& \delta_p(t) 						\\
\end{bmatrix}
\end{equation}
The Rabi frequencies $\Omega_m(t)$ 
are related to the amplitudes of the pump and Stokes fields by 
$\Omega_s = Q_{12} \,{\cal A}_s(t)$ and $\Omega_p = {Q}_{02} \,{\cal A}_p(t)$.
The single-photon detunings are $\delta_s=\epsilon_2-\epsilon_1-\omega_s$ and 
$\delta_p=\epsilon_2-\epsilon_0-\omega_p$, whereas 
$\delta = \delta_p - \delta_s$ is the two-photon detuning.  
At a fixed time and at two-photon resonance, $\delta=0$, 
the Hamiltonian (\ref{eq:H}) has an eigenvector 
with null eigenvalue, $E_D=0$
\begin{equation}
\label{eq:dark.state}
\ket{D}=\frac{\Omega_s \ket{0} - \Omega_p \ket{1}}{\sqrt{\Omega_s^2+\Omega_p^2}}
\end{equation}
which is called the ``dark state'' since 
despite of the fields triggering transitions 
$\ket{2}$ is depopulated. Indeed if the system 
in state $\ket{D}$ destructive interference cancels absorption 
and there is no subsequent atomic fluorescence. 
The other eigenstates $\ket{\pm}$ have eigenvalues $E_{\pm}= \frac{1}{2}\delta_p \pm \frac{1}{2} \Omega_{AT}$ where 
$\Omega_{AT} =
\sqrt{\Omega_p^2 + \Omega_s^2 + \delta_p^2}$ is the 
Autler Townes splitting. 

Adiabatically time-dependent Rabi couplings $\Omega_m(t)$  
allow to span the trapped subspace 
by evolving the 
instantaneous $\ket{D(t)}$. This is the principle of 
ideal STIRAP: two pulses shined in the 
so called \textit{counterintuitive} sequence, i.e. the Stokes pulse is shined before the 
pump one
(Fig.~\ref{fig:stirap-ideal}b)  transform 
$\ket{D(t)}: \ket{0} \to \ket{1}$, corresponding to complete
population transfer (Fig.~\ref{fig:stirap-ideal}d). 
As an example Gaussian-shaped pulses 
we use are given  by
\begin{equation}
\label{eq:pulses}
\Omega_p= \kappa_p \Omega_0 \,\mathrm{e}^{-[({t+\tau})/{T}]^2} \;\;;\;\, 
\Omega_s=\Omega_0 \,\mathrm{e}^{-[({t-\tau}/{T}]^2}; 
\end{equation}

Adiabaticity is usually guaranteed by strong enough drive, $\Omega_0 T \gtrsim 10$.  
STIRAP involves several coherent phenomena~\cite{kr:201-vitanov-advatmolopt}, thereby it is a benchmark for multilevel advanced control. Besides allowing $\sim 100\%$ efficiency of coherent transfer, 
its success lies in the striking insensitivity to small variations of control parameters. 
Actually parametric fluctuations of $\delta$ are the most detrimental, since for $\delta \neq 0$ 
there is no dark state implementing the adiabatic linkage $\ket{0}\to \ket{1}$, but still STIRAP 
occurs via Landau-Zener processes~\cite{kr:201-vitanov-advatmolopt}. Concerning to decoherence 
processes involving the intermediate state $\ket{2}$ are not relevant, while dephasing in the 
"qubit" subspace $\mathrm{span}\{\ket{0},\ket{1}\}$ is the most detrimental ~\cite{ka:213-falci-prb-stirapcpb}.

In AAs we can obtain STIRAP by applying a two-tone electric or magnetic field 
at $\mathrm{GHz}$ frequencies. The coupling operator $Q$ depends on the design, being 
the charge operator in devices of the transmon family~\cite{ka:207-koch-pra-transmon,ka:212-rigettisteffen-prb-trasmonshapphire} 
or the loop current in the flux-based devices~\cite{ka:200-walmooij-science-superposition,ka:211-bylander-natphys}. In such devices the dephasing processes which are in principle the most detrimental   
are due to low-frequency noise sources coupled via the operator $Q$. They induce fluctuations of the qubit energies, which in turn translate in fluctuations of detunings in the Hamiltonian (\ref{eq:H}). 
They are suppressed In last-generation devices, displaying decoherence times up to $\sim 0.1\,\mathrm{ms}$, by tuning the static component of the external fields in such a way 
that the AA Hamiltonian $H_0$ assumes a symmetric form. Effects of fluctuations vanish in 
lowest order, together with many matrix elements, ${Q}_{ii}={Q}_{02}= 0$. Unfortunately 
this last equality expresses a selection rule preventing the pump pulse to be operated, since it yields $\Omega_p=0$ in Eq. (\ref{eq:H}). This reduced available control constraints has to be circumvented 
in order to implement STIRAP in last-generation devices.


\section{New control schemes for STIRAP in artificial atoms}
\label{sec:always-on}
\subsection{2+1 STIRAP: always-on pump coupling at symmetry points}
\label{sec:STIRAP21}
These difficulties raise the question on whether or not it is possible to find a control solution that gives rise to the same physics as the STIRAP taking advantage of the "ladder" couplings alone. We found that this is indeed possible by employing a two-photon pump pulse supplemented 
by a suitable phase-modulation control~\cite{ka:216-distefano-pra-twoplusone}. Here we study a version 
which uses reduced control resources, allowing for one of the couplings being always-on, 
We consider a two tone pump field ${\cal A}_p(t) = {\cal A}_{p1}(t) \cos \omega_{p1} t + {\cal A}_{p2}\, \cos \omega_{p2} t$ where $\omega_{p1}\simeq E_1$ and $\omega_p2 \simeq E2-E_1$. The RWA Hamiltonian for a three-level system coupled to this field can be transformed in a suitable doubly rotated frame to yield the standard form
\begin{equation}
\label{eq:2pump-H}
H_p = H = \begin{bmatrix}
0 								& 	\frac{1}{2}\Omega_{p1} 							& 0	\\
\frac{1}{2}\Omega_{p1} 								& 	\delta_2				& \frac{1}{2}\Omega_{p2} 	\\
0 		& 	\frac{1}{2}\Omega_{p2} 	& \delta_p						\\
\end{bmatrix}
\end{equation}
Where $\delta_2 := E_1-\omega_{p1}, \delta_p:=-(\omega_{p1} + \omega_{p2})$ and we introduced the Rabi frequencies $\Omega_{p1}:=Q_{01}{\cal A}_{p1}, \Omega_{p2}:=Q_{12}{\cal A}_{p2}$. If the pump fields are well detuned from one-photon resonance, $\Omega_k\ll \delta_2, |\delta_2-\delta_p|$, 
state $\ket{1}$ can be adiabatically eliminated, yielding an effective Hamiltonian
capturing the coarse-grained dynamics over a time scale of $\Delta t : |\Omega_k|\ll 1/\Delta t \ll |E_i-E_j|$. 
\begin{equation}
\label{eq:ave-H-p}
H_{ave}^{\prime} = \begin{bmatrix}
0 								& 	0 							& \frac{1}{2}\Omega_p	\\
0 								& 	\delta_2	-(2S_1-S_2)			& 0 	\\
\frac{1}{2}\Omega_p 		& 	0 	& \delta_p 						\\
\end{bmatrix}
\end{equation}
where we also take into account the Stark shifts induced by the two-photon drive in second order perturbation theory
\begin{equation}
\label{eq:par_ave-H}
\Omega_p = -\frac{\Omega_{p1}\Omega_{p2}}{2\delta_2}\;;
\quad S_1 = - \frac{|\Omega_{p1}|^2}{4\delta_2}\,;
\quad S_2 = - \frac{|\Omega_{p2}|^2}{4(\delta_2-\delta_p)}
\end{equation}
In order to perform STIRAP we add a suitably modulated Stokes field ${\cal A}_s(t) \cos [\omega_s t +\phi(t)]$. In the rotated frame of Eq.(\ref{eq:2pump-H}) the corotating part of the 
control term then reads $H_S =\Omega_s \mathrm{e}^{i[(\delta-\delta_2)t +\phi(t)]} \ket{1}\bra{2}+
\mbox{h.c.}$, where $\Omega_s(t):=Q_{12}{\cal A}_s(t)$. If we further   
rotate the Hamiltonian by 
$U_s = 
\mathrm{e}^{i[(\delta-\delta_2)t +\phi(t)]\ket{1}\bra{1} }$
we obtain finally 
\begin{equation}
\label{eq:ave-H}
H_{ave} = \begin{bmatrix}
0 								& 	0 							& \frac{1}{2}\Omega_p	\\
0 								& 	\delta	-(2S_1-S_2) + \dot{\phi}(t)			& \frac{1}{2} \Omega_s 	\\
\frac{1}{2}\Omega_p 		& 	\frac{1}{2} \Omega_s 	& \delta_p 						\\
\end{bmatrix}
\end{equation}
This result is obtained in a most rigorous way in the framework of the Average Hamiltonian Theory. 
Now, if we choose $\dot{\phi}(t) = 2S_1-S_2$, the effective Hamiltonian 
reduces to of Eq.(\ref{eq:H}), and may implement STIRAP.
Therefore we expect that the full Hamiltonian implements STIRAP  even 
in the absence of a direct pump coupling, provided 
pulses $\Omega_s(t)$ and $\Omega_{pk}(t)$ are shined in the counterintuitive sequence, 
this Hamiltonian implements STIRAP in the absence of a direct pump coupling and $\delta$ is small
enough to ensure nearly two-photon resonance. This control scheme allows to observe STIRAP in last-generation superconducting AAs~\cite{ka:216-distefano-pra-twoplusone}.

In principle the same conclusio holds also if we impose a further constraint, namely one of the 
components of the two-photon pump fields is always on.  
\begin{figure}[t!]
\centering
\includegraphics[width=0.3\columnwidth]{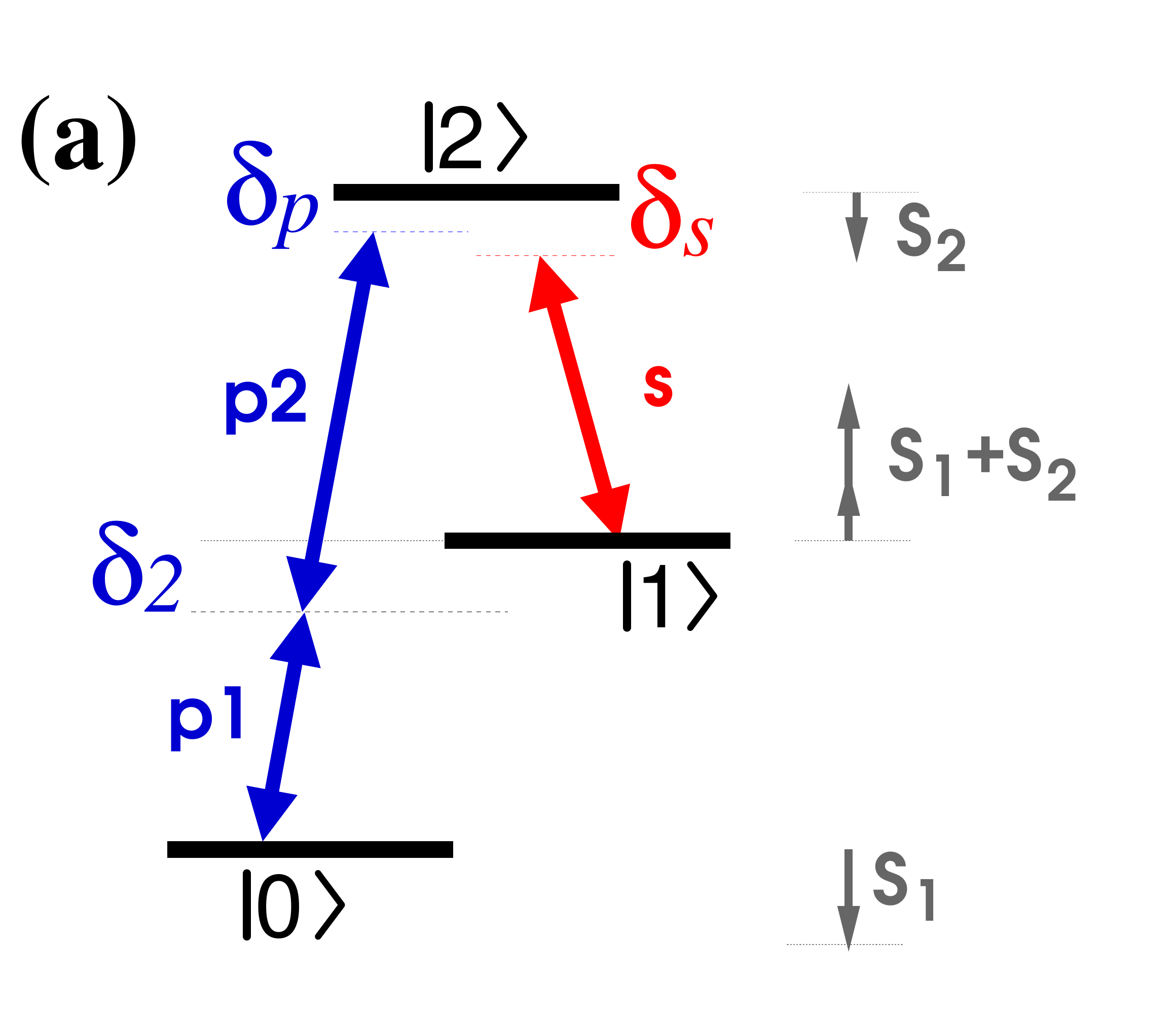}
\qquad
\includegraphics[width=0.4\columnwidth]{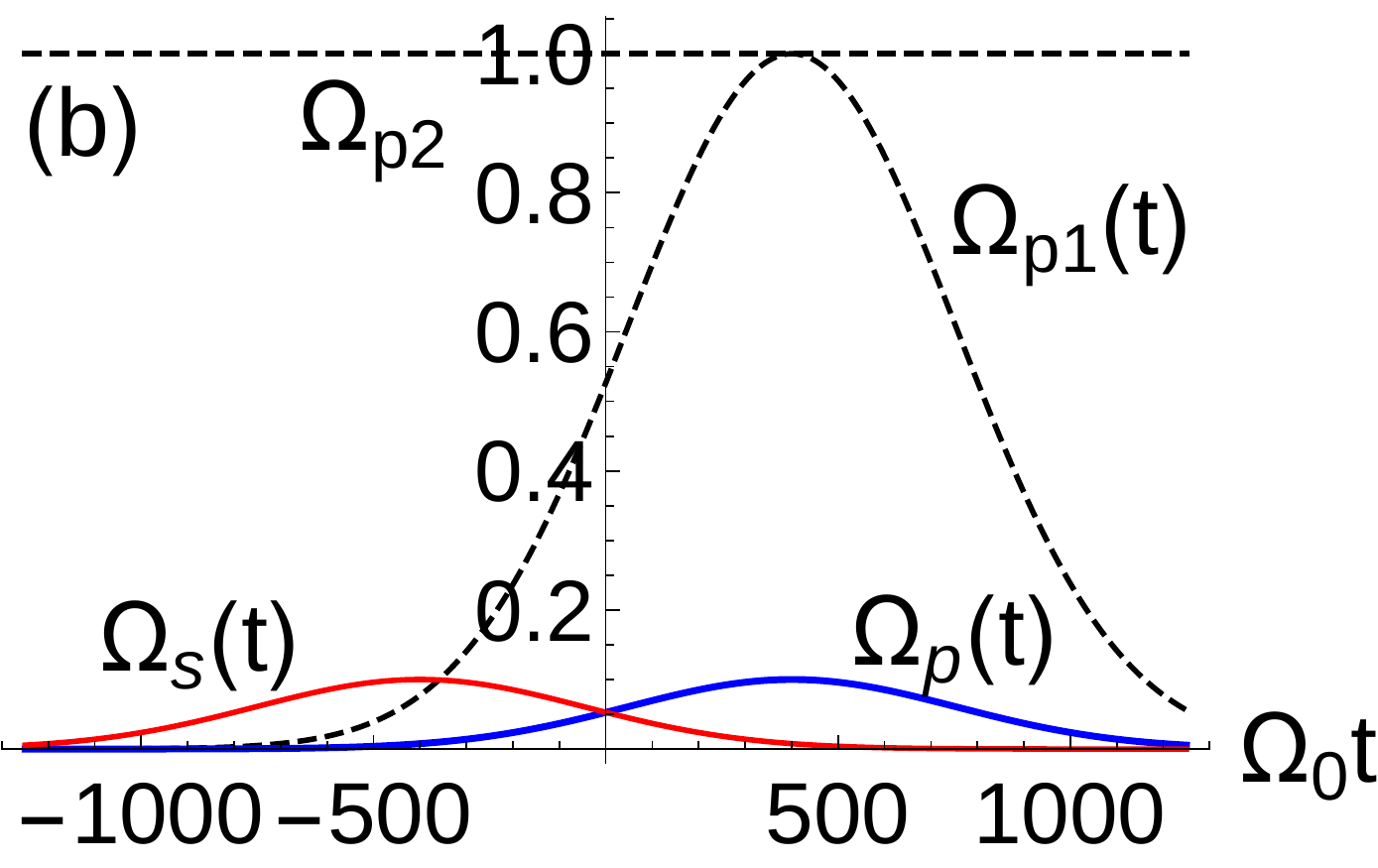}
\\
\includegraphics[width=0.49\columnwidth]{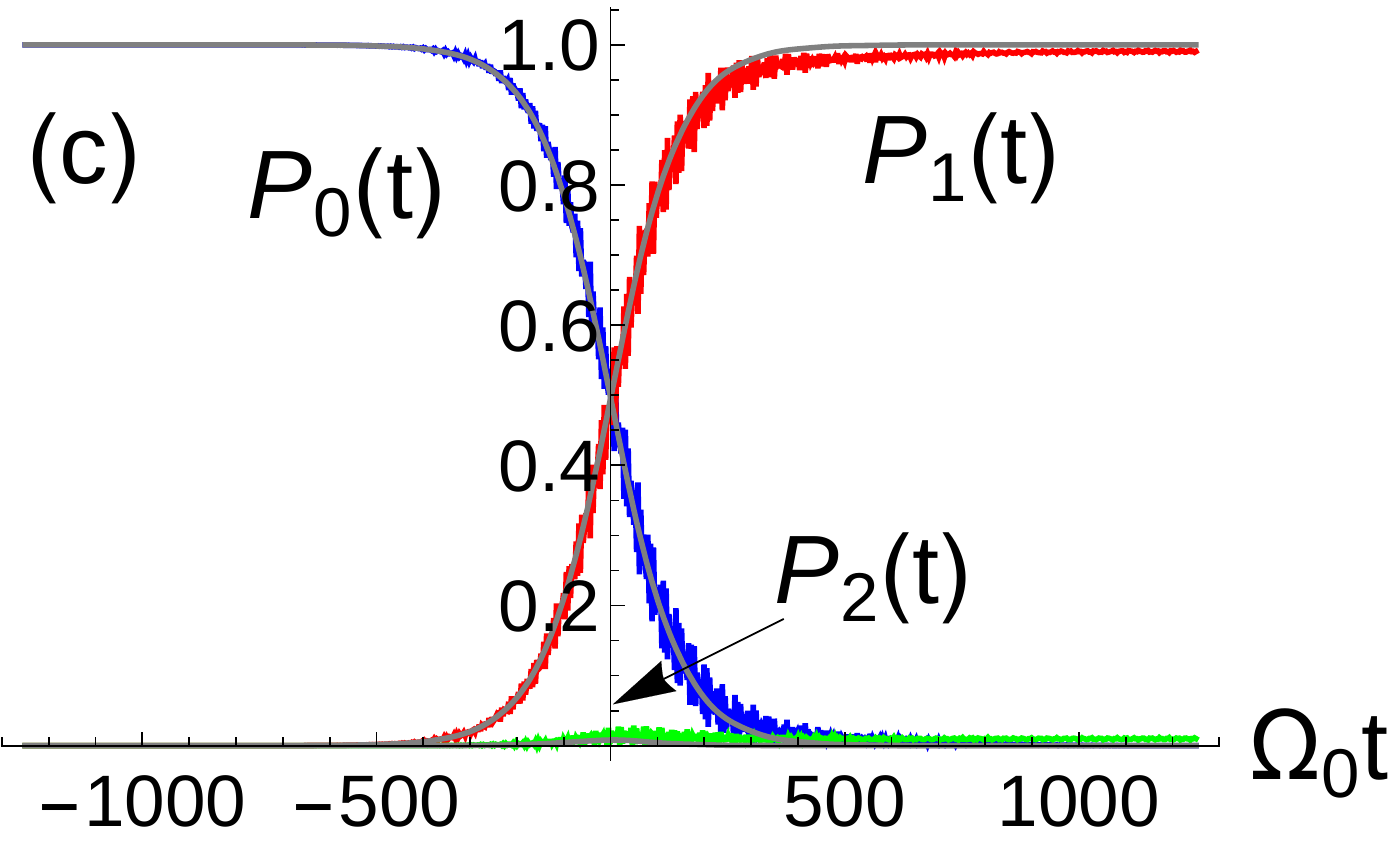}
\hfill
\includegraphics[width=0.49\columnwidth]{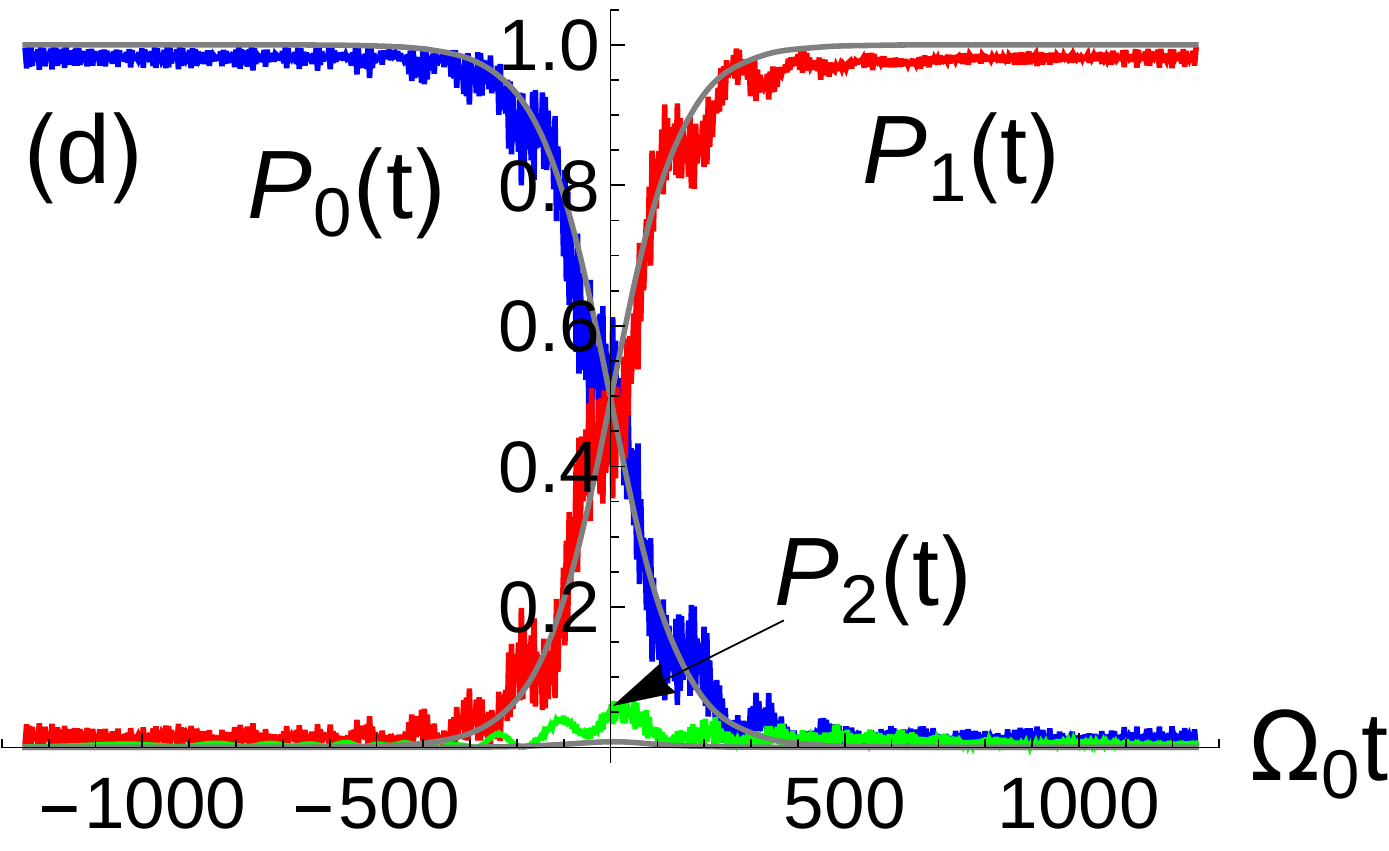}
\caption{
(a) couplings of the $\Lambda$ system with a 2-photon pump pulse, determining the Stark shifts $S_1$ and $S_2$; (b) External and effective fields in units of $\Omega_0$; the detuned ($\delta_2/\Omega_0=5$) constant drive 
$\Omega_{p2}$ and the Gaussian $\Omega_{p1}(t)$ yield an effective $\Omega_p(t)$ with amplitude
much smaller than $\Omega_0$; $\Omega_s(t)$ is chosen with the same amplitude to maximize the efficiency. 
(c) Populations histories vs $\Omega_0 t$ of the 2+1 STIRAP protocol (coloured lines), for 
always-on $\Omega_{p2}$. Gray lines are the populations of the effective Hamiltonian Eq. (\ref{eq:ave-H}), capturing very well the coarse-grained dynamics. 
Here 
$\mathrm{Max}[\Omega_{s,p}(t)] T = 50$
(d) 
The protocol with always-on $\Omega_{p1}$ with the same figures, proves to be less faithful.
\label{fig:stirap21}
}
\end{figure}
Fig.~\ref{fig:stirap21}(c) shows that the results of the numerical simulations for 
always on $\Omega_{p2}$ which agree very well with the theoretical arguments given above. Here the relevant figure of merit for adiabaticity is estimated from the effective Hamiltonian. 
If we choose $\Omega_s(t)$ with the same amplitude of the effective $\Omega_p(t)$, which 
maximizes the efficiency, good adiabaticity is obtained if 
$\mathrm{Max}[\Omega_{s,p}(t)] T \gg 10$~\cite{kr:198-bergmann-rmp-stirap}.
It is also seen that the effective Hamiltonian obtained by the Average Hamiltonian Theory reproduces very well the coarse-grained dynamics of STIRAP. Therefore this protocol allows work at symmetry points
of the AA Hamiltonian exploiting the advantages of noise protection~\cite{ka:216-distefano-pra-twoplusone}. Notice that the state $\ket{1}$ is at the same time the virtual intermediate level for the two-photon pump and the real target level for STIRAP. 

We notice that, while the effective Hamiltonian is invariant if we change $p1 \leftrightarrow p2$, 
protocols with always-on $\Omega_{p1}$ are less faithful, since the constant field produces small 
Rabi oscillations between $\ket{0}$ and $\ket{1}$ at the beginning of the protocol, when $\ket{0}$ 
is populated. Instead the always-on $\Omega_{p1}$ scheme works better for 
the time-reversed protocol.


\subsection{c-STIRAP: implementing control by detunings}
\label{sec:cSTIRAP}
In this section, we discuss a protocol to achieve coherent population transfer with 
an always-on Stokes field, mainly operating on detunings.
In the system Hamiltonian Eq.(\ref{eq:H}) we take $\Omega_s(t) = \Omega_0$. In this case the dark state Eq.(\ref{eq:dark.state}) does not implement population transfer from $\ket{0}$ to $\ket{1}$ under the two-photon resonance condition, thereby fields must be effectively switched by a modulation of 
detunings. In addition, in the presence of the always-on field, the target state $\ket{1}$ is no more  eigenstate of the Hamiltonian at the end o the protocol. We shall then shape the detunings in a way that asymptotically $\delta_s(t\to\pm\infty)\gg\Omega_0$.
\begin{figure}
\centering
\includegraphics[width=0.49\columnwidth]{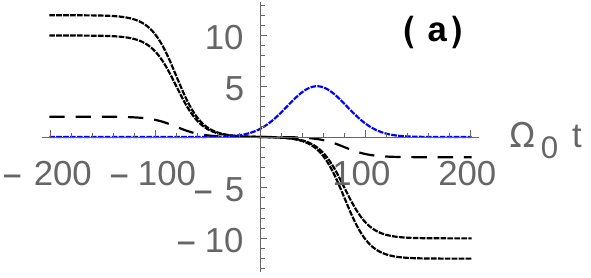}
\includegraphics[width=0.49\columnwidth]{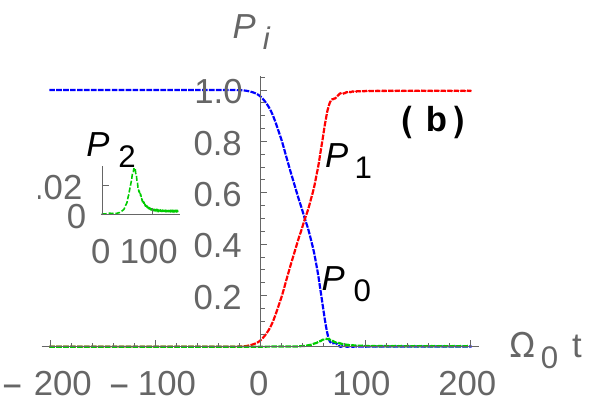}
\caption{
(a) single-photon (black solid lines) and two-photon (black dashed line) detunings of the cSTIRAP protocol, together with a 5 times magnified pump pulse (blue line); (b) populations for the cSTIRAP protocol with $\Omega_s(t) = \Omega_0, \Omega_0T=40, h_{\delta}=10, \kappa_{\delta}=1.2$ and $\kappa= 1$, $\tau_{ch}=0.6T$. In the inset a magnification of the transient population of $\ket{2}$ is displayed, showing that it can be kept negligibly small during the whole duration of the protocol.
\label{fig:cstirap}
}
\end{figure}
Our proposal is to employ detunings shaped as in Fig.~\ref{fig:cstirap}(a)
\begin{equation}
\label{eq:detunings-cstirap}
\begin{aligned}
\delta_s(t) &= \frac{1}{2} \, 
h_{\delta} \Omega_0 \left[\tanh\left(\frac{t-\tau}{\tau_{ch}}\right) + \tanh\left(\frac{t+\tau}{\tau_{ch}}\right)\right] \\
\delta_p(t) &= \kappa_{\delta} \delta_s(t)
\end{aligned}
\end{equation}
and a Gaussian pump field $\Omega_p = \kappa_p \, \Omega_0 \mathrm{e}^{-\left(\frac{t-t_c}{T}\right)^2}$. The main feature of the detuning modulation of Fig.~\ref{fig:cstirap} (a) is that, while detunings are big at the beginning and the end of the protocol ($h_{\delta} \gg 1$), a resonant stage holds in the central part of the protocol. It is in the resonant stage, when $\delta=0$, that the pump pulse is turned on, allowing for some population transfer via the STIRAP mechanism. Then, at about the maximum of the pump pulse, the transfer is completed by detuning the fields and eventually the pump pulse is turned off. In fig. \ref{fig:cstirap} (b) we show the population histories for $\Omega_0 T = 40$ and $\kappa=1$. It is seen that, through the interplay between the resonant and far detuned stage, the transient population of state $\ket{2}$ can be kept considerably small. Further details on the working principles and the analysis of the stability of the protocol are beyond the scopes of the present paper have been thoroughly discussed in Ref.~\cite{ka:215-distefano-prb-cstirap}. Here we want to point out that we found the protocol to be stable against static deviation of the parameter, the most critical one being the two-photon detuning. It is therefore clear that most of the arguments given above for the STIRAP apply also to this new protocol.

\section{STIRAP in circuit QED architectures}
\label{sec:circuitQED}
In this section we show how substituting the always-on fields with quantized cavity modes 
the protocols introduced in last sections may perform tasks in circuit-QED architectures, where 
coupling between AAs and microwave cavities is not directly switchable. 
The basic building block of circuit-QED 
is a two-level atom (states $\ket{g}$ and $\ket{e}$) with Bohr splitting $\varepsilon$ coupled to an electromagnetic mode with angular frequency $\omega_c$, described by the
Jaynes-Cummings model of quantum optics
\begin{equation}
\label{eq:JC2}
\begin{aligned}
H_{JC2} &=  \varepsilon \, \ketbra{e}{e} + \omega_c  \,a^\dagger a
+ g \big(a^\dagger \ketbra{g}{e} + a \ketbra{e}{g}\big)
\end{aligned}
\end{equation}
The quantized field is coupled in the RWA, via the coupling constant $g$. The RWA 
is routinely used in many different setups, 
where couplings are smaller than $\sim 1\%$ of 
$\omega_c$ and $\varepsilon$, as superconducting AAs~\cite{ka:204-wallraff-nature-cqed}. It describes light-matter 
interaction when the coupling 
overcomes decoherence rates of cavities and atoms, $\kappa,\gamma \ll g$,
the so called strong coupling regime. The ground state of this Hamiltonian is 
factorized in the cavity vacuum and the atomic ground stare, 
$\ket{0g} := \ket{0} \otimes \ket{g}$,  
whereas the other eigenvectors can be found in the two-dimensional invariant subspaces 
$\mathbbm{H}^{(n)}=\mathrm{span}\{ \ket{n-1 \, e}, \ket{n\,g }\}$.  
To implement a $\Lambda$ scheme we first introduce an auxiliary 
atomic level $\ket{b}$ of lower energy $- \varepsilon_b < 0$. The resulting Hamiltonian is (see Fig.~\ref{fig:usc1}a)
\begin{equation}
\label{eq:JCH3}
\begin{aligned}
H_0 = -\varepsilon_b \,\ketbra{b}{b} + H_{JC} + \omega_c \,a^\dagger a \otimes \ketbra{b}{b}
\end{aligned}
\end{equation} 
where we assume that the auxiliary level is sufficiently detuned from the cavity 
$|\varepsilon_b- \omega_c|  \gg g$ that further terms in the interaction can be neglected.

\subsection{Implementing control via detunings}
We now seek to adapt the c-STIRAP protocol of 
\S\ref{sec:cSTIRAP}  to the circuit-QED architecture. To this end we 
consider a control pump field of the form 
$H_c(t)= \Omega_p(t) \,\cos (\omega_p t) \,\big[ \ketbra{e}{b} + \ketbra{b}{e} \big]$
Since it couples each $\ket{n}\otimes\ket{b}$, for $n=0,1,\dots$ only with states of the invariant subspace $\mathbbm{H}^{(n+1)}$, the dynamics factorizes in sectors of three-level 
subspaces. In particular for $n=0$ the dynamics is governed by the projected Hamiltonian, which 
in the basis $\{ \ket{0b}, \ket{1g}, \ket{0e}\}$ reads  
\begin{equation}
\label{eq:cstirap-mode-labframe}
H = 
\left[
\begin{array}{ccc}
-\varepsilon_b & 0 & \Omega_p(t) \cos (\omega_p t)
\\[2pt]
0 & \omega_c & g 
\\[2pt]
\Omega_p(t) \cos (\omega_p t)& g & \varepsilon  
\end{array}
\right]
\end{equation}
Letting $\varepsilon + \varepsilon_b -\omega_p = \delta_p(t)$,  
$\varepsilon -\omega_c = \delta_s(t)$ and $\delta(t)= \delta_p(t) - \delta_s(t)$, neglecting 
counterrotating terms and performing a transformation to a rotating frame defined by 
$U_x(t)= \mathrm{e}^{i \omega_p t \ketbra{0b}{0b}}$ we finally pervent to the effective Hamiltonian
$\tilde{H} =  U_x^\dagger \, H \, U_x - i\,U^\dagger_x \, (\partial_t U_x)$. It has the 
same form of Eq.(\ref{eq:H}), apart from the substitution $\Omega_s(t) \to 2 g$. Therefore the 
coupling to a quantized cavity mode acts as an always-on Stokes field in the reduced dynamics.
In particular operating the c-stirap  protocol of \S\ref{sec:cSTIRAP} would yield population 
transfer $\ket{0b}  \to \ket{1g}$, thereby a single-photon is injected in the cavity. 
Then this procedure implements vacuum stimulated Raman adiabatic passage
(v-STIRAP)~\cite{ka:202-kuhnetal-prl-singleph} with the constraint of 
an always on coupling to the cavity. 

If $\gamma \gg  \kappa$ the target state will decay $\ket{1g} \to \ket{1b}$ and 
a second photon can be injected via a STIRAP cycle in the subspace spanned by 
$\{ \ket{1b}, \ket{2g}, \ket{1e}\}$. Notice that in this case the Hamiltonian has the form   
(\ref{eq:cstirap-mode-labframe}) with $g \to \sqrt{2} g$. This does not require any modification
of the timing of the protocol, since STIRAP is rather insensitive to the precise amplitude of the 
driving fields. Repeating the protocol would produce a Fock state in the cavity. 
We stress this protocol can be adapted to AAs where selection rules prevent to couple 
directly the pump field by using a two-photon pump as in \S\ref{sec:STIRAP21}. 
In this case no field modulation is needed since the intermediate state, e.g.  
$\ket{0g}$ for a single photon injection, is not involved in the dynamics, then acting 
as a fourth auxiliary level in standard multiphoton STIRAP~\cite{ka:205-fewell-optcomm-adiabelimin}

Finally while the c-stirap protocol we illustrated involves the emission of a single photon 
to the cavity, it is possible to design a dual protocol with always-on quantized pump field 
and in this case a cavity photon will be converted in an atomic excitation, leading in principle 
to cooling when several cycles are considered. 

\subsection{Implementing control via a trigger field}
Now we turn to the 2+1protocol of \S\ref{sec:STIRAP21}. 
Substituting the always-on field with a cavity, the STIRAP process we studied involves the absorption of a single photon, while the inverse cycle determines photon emission. 
This can be shown by consider a two-tone control field, which for proper choices of 
frequencies and not too large amplitudes reduces to the effective form 
$H_c(t)= \Omega_{p1}(t) \,\cos (\omega_{p1} t) \,\big[ \ketbra{b}{g} + \ketbra{g}{b} \big]+ 
\Omega_{s}(t) \,\cos (\omega_{s} t) \,\big[ \ketbra{g}{e} + \ketbra{e}{g} \big]
$, where 
$\omega_{p1}= \varepsilon_b - \delta_2$, $\omega_c = \varepsilon + \delta_2$ 
and $\omega_s=  \varepsilon$.
As in the last section the Hamiltonian expressed in the product basis allows to restrict the 
dynamics to the subspace. The procedure to pervent to an effective STIRAP Hamiltonian is more complicated and for clarity here we outline it neglecting dispersive coupling of fields 
not relevant in two-photon processes, i.e. the two-photon pump and the dynamics of the 
$\Lambda$ configuration needed for STIRAP. In this approximation 
we can consider the subspace $\mathrm{span}\{\ket{0g},\ket{0e},\ket{1b},\ket{1g},\ket{1e}\}$.
Performing the RWA and transforming in a suitable triple 
rotating frame we find the following 
truncated Hamiltonian
$$H_{5}
= \left[
\begin{array}{ccccc}
0 & \Omega_s/2 & 
\\
\Omega_s/2 & 0 & 0 &g &
\\ &0&0 &\Omega_{p1}/2 & \\
 &g&\Omega_{p1}/2 & \delta_2 & \Omega_s/2
 \\
 &&&  \Omega_s/2 & \delta_2
\end{array}
\right] 
$$ 
where we dropped irrelevat zeroes in the matrix. For $\delta_2 \gg g, \Omega_{p1}$ the central 
$3\times 3$ block can be approximated by an effective Hamiltonian, which overall yields
$$
H_{5eff}
= \left[
\begin{array}{ccccc}
0 & \Omega_s/2 & 0 
\\
\Omega_s/2 & S_g & \tilde{g} & 0 &
\\0  &\tilde{g} & S_{p1} & 0 \\
 &0&0 & \delta_2 - (S_g+S_{p1}) & \Omega_s/2
 \\
 &&&  \Omega_s/2 & \delta_2
\end{array}
\right] 
$$ 
where 
\begin{equation}
\label{eq:effcoupling}
S_g = {g^2 \over \delta_2}	\;,\quad S_{p1}(t)= - {\Omega_{p1}^2(t) \over 4 \delta_2}\;,\quad 
\tilde{g}(t) = - {g\Omega_{p1}(t)\over 2 \delta_2}
\end{equation}
The upper-left $3\times 3$ block implements 
a $\Lambda$ scheme potentially yielding STIRAP between $\ket{0g} \leftrightarrow  \ket{1b}$, 
depending on the preparation and the pulse sequence. Atom-cavity exchange of excitations  
occurs thanks to the fact that the external field $\Omega_{p1}(t)$ activates  
the interaction $g$, thus implementing an effective switchable coupling $\tilde{g}(t)$
given by Eq.(\ref{eq:effcoupling}). This is a generalization of the switchable coupling 
scheme recently introduced in Ref.~\cite{ka:215-zeytinogluwallraff-phasetunable}. 
In our case we must cancel the potentially dangerous dynamical Stark shift $S_{p1}(t)$,  
which can be done by phase 
modulation of the control fields
as explained in \S\ref{sec:STIRAP21}, whereas $S_0$ is cancelled by proper static 
detunings.

\section{Probing ultrastrong coupling by STIRAP}
\label{sec:USC}
\begin{figure}[t!]
\resizebox{43mm}{40mm}{\includegraphics{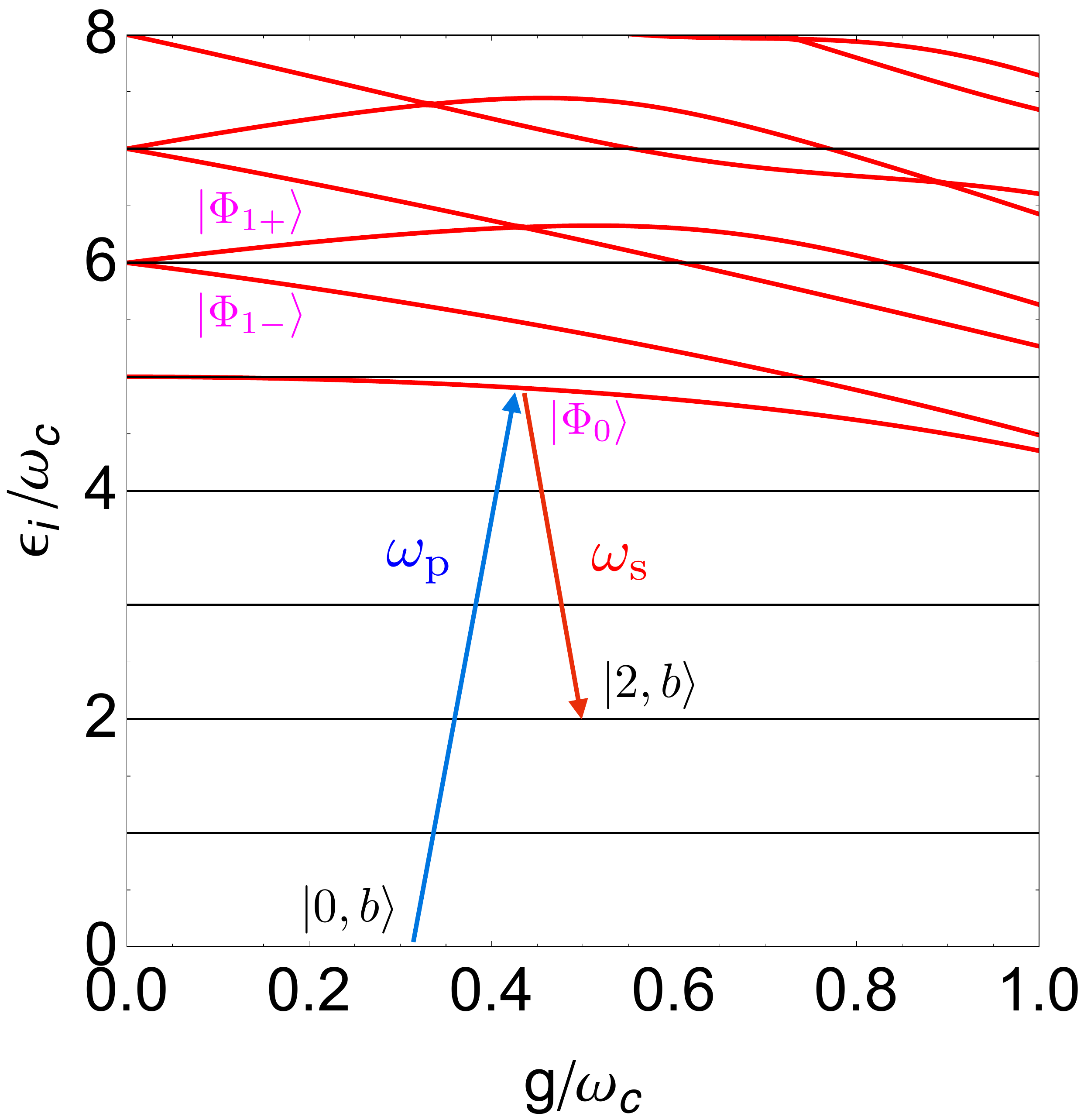}}
\resizebox{43mm}{40mm}{\includegraphics{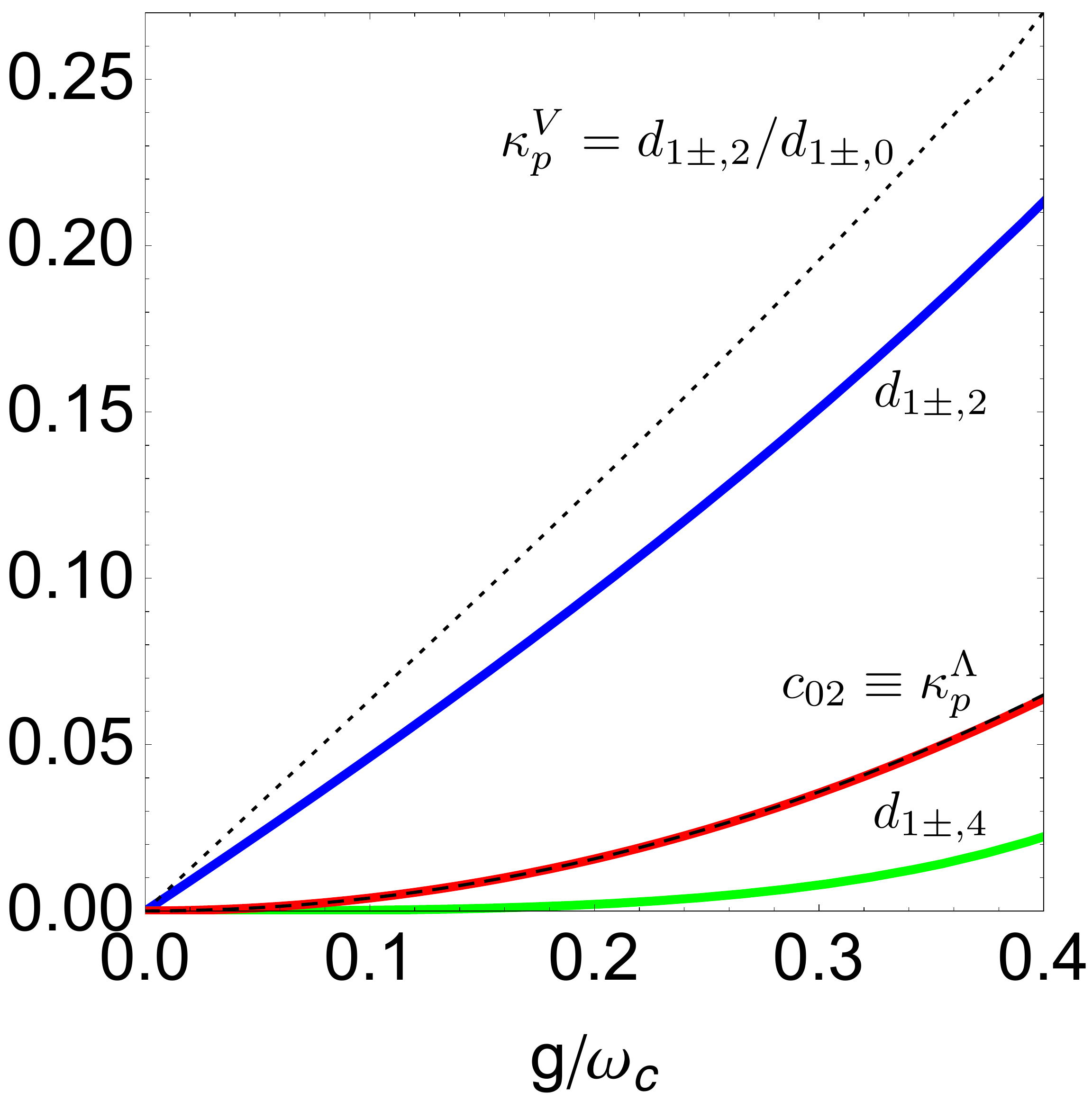}}
\caption{
(a) Spectrum $\{E_i\}$ of dressed states of a three-level atom coupled to a cavity, from strong coupling to the USC regime. Here
$\varepsilon_b= 5 \omega_c$ and  $\varepsilon=\omega_c$.
Thin black lines correspond to unmixed $\ket{n\,b}$ states, whereas red ticker lines are the eigenstates of the Rabi Hamiltonian Eq.(\ref{eq:rabiH2}): for small $g/\omega_c$ they are 
linear in $g$, as in the Jaynes-Cummings model, deviations from linearity being an effect of USC.   
(b) Amplitudes of the eigenstates $\ket{\Phi_j}$ of the Rabi model reducing to zero in the 
Jaynes-Cummings limit. For $g/\omega_c$ not too large the relevant ones for STIRAP 
are $c_{02}(g)$ for $\ket{\Phi_0}$, and $d_{1\pm,n}=\braket{n\,e}{\Phi_{1\pm}}$. The 
quantities $\kappa_p^{\Lambda,V}$ represent the optimal attenuation of the external 
pump field, yielding the best efficiency and robustness performances of the protocol for STIRAP 
via $\ket{\Phi_0}$ and $\ket{\Phi_{1 \pm}}$. 
\label{fig:usc1}}
\end{figure}

An interesting application of STIRAP is coherent amplification of signature of 
ultrastrong coupling (USC) in AAs coupled to solid-state cavities. 
The physics is captured by the Rabi Hamiltonian describing the full dipole coupling
of a two-level atom to an electromagnetic mode 
\begin{equation}
\label{eq:rabiH2}
\begin{aligned}
H_{R2} &=  H_{JC2} + g  \big(a \ketbra{g}{e} + a^\dagger \ketbra{e}{g}\big)
\end{aligned}
\end{equation}
The last "counterrotating" term in $H_{R2}$ must be taken in to account when $g$ increases.
Nanofabrication offers the possibility of producing architectures of AAs
and quantized modes with 
$g/\omega_c \sim 0.1-1$, entering 
the regime of ultrastrong coupling (USC), where light and matter experience still unexplored non-perturbative phenomena. 
Experiments performed so far have 
detected spectroscopic features of USC~\cite{ka:210-niemczyck-natphys-ultrastrong,ka:212-scalari-science-USCTHz},
as the Bloch-Siegert shift in the energy spectrum (\ref{fig:usc1}a). 
Proposals of dynamical detection of 
USC rely on the fundamental property that, contrary to the Jaynes-Cummings ground state $\ket{0g}$, 
the Rabi ground state $\ket{\Phi_0}$ contains photons. In particular it has the structure 
$\ket{\Phi_0} = \sum_{n=0}^\infty c_{0\,n} \ket{ng} + d_{0n} \ket{n\,e}$  (see Fig.~\ref{fig:usc1}b)
with the restriction that only an even number of excitions appears. One introduces an auxiliary 
atomic level $\ket{b}$ of lower energy $- \varepsilon_b$. The resulting Hamiltonian is (see Fig.~\ref{fig:usc1}a)
\begin{equation}
\label{eq:rabiH3}
\begin{aligned}
H_0 = -\varepsilon_b \,\ketbra{b}{b} + H_{R2} + \omega_c \,a^\dagger a \otimes \ketbra{b}{b}
\end{aligned}
\end{equation} 
provided the auxiliary level is far detuned from the cavity $|\varepsilon_b| \gg \omega_c$. 

In Ref.~\cite{ka:213-stassisavasta-prl-USCSEP} detection of USC by spontaneous emission pumping (SEP) was proposed: population pumped from $\ket{0b}$ to $\ket{\Phi_0}$
may decay in $\ket{2b}$,  due to the finite overlap 
$c_{02} = \braket{2g}{\Phi_0} \neq 0$. The process leaves two-photons in the cavity, 
detection of this channel being an unambiguous signature of USC. In another 
proposal~\cite{ka:214-huanglaw-pra-uscraman} the system 
is driven by a two-tone classical electromagnetic field addressing mainly the two lowest 
atomic levels 
$$
\begin{aligned}
H_c(t) &= W(t) \,(\ketbra{b}{g}+\ketbra{g}{b}) \\
&= 
W(t) \sum_{nj} c_{jn} [\ketbra{n\,b}{\Phi_j}+\ketbra{\Phi_j}{n\,b}].
\end{aligned}
$$ 
where $W(t)=\mathscr{W}_s(t) \,\cos \omega_s t +\mathscr{W}_p(t) \,\cos \omega_p t$ and
$\mathscr{W}_k(t)$ with $k=p,s$ are slowly varying envelopes fields. We choose 
$\omega_p \approx \varepsilon_b + E_0 - \delta_p$  and 
$\omega_s \approx \varepsilon_b - 2 \omega_c + E_0 - \delta_p$. For large detuning  
$\delta_p \gg  \max[\mathscr{W}_k(t)]$ pulses with proper timing produce a Raman oscillation  
transforming $\ket{0b} \to \ket{2b}$. Again this is possible only if 
$c_{02} \neq 0$. Population transfer is also possible with suitably crafted resonant 
pulses. Here we show that a Hamiltonian with the same structure allows to implement the dynamical detection of USC via a STIRAP protocol, which is expected to be efficient and robust.
Comparison between the proposals is discussed in  
to \S\ref{sec:conclusions}. 
The essential physics is well captured in the limit of not too strong fields $\mathscr{W}_k$ and coupling $g$. Then $H_c$ greatly simplifies, yielding 
$$
H_c^\Lambda(t)  = \frac{\Omega_{p}(t)}{2} \mathrm{e}^{i \omega_p t} \ketbra{0\,b}{\Phi_0}+
\frac{\Omega_{s}(t)}{2} \mathrm{e}^{i \omega_s t} \ketbra{2\,b}{\Phi_0} + 
\mbox{h.c.}
$$
where $\Omega_p=c_{00}(g) \mathscr{W}_p$ and $\Omega_s=c_{02}(g) \mathscr{W}_s$. 
This is a $\Lambda$ configuration~\cite{kr:198-bergmann-rmp-stirap,kr:201-vitanov-advatmolopt}, 
the relevant dynamics being restricted to three levels. If we truncate $H_0 \approx -\varepsilon_b \ketbra{0b}{0b} - (\varepsilon_b-2 \omega_c) \ketbra{2b}{2b} + E_0  \ketbra{\Phi_0}{\Phi_0}$, 
we obtain the Hamiltonian $H=H_0+H_c^\Lambda(t)$ for STIRAP\cite{kr:198-bergmann-rmp-stirap,kr:201-vitanov-advatmolopt}. For STIRAP we prepare the system in  $\ket{0b}$ and shining the two pulses of width $T$ in the counterintuitive sequence. We expect $~\sim 100\%$ population transfer to $\ket{2b}$ if adiabaticity is sufficient, i.e. using large pulse areas $\max_t [\Omega_k(t)] T > 10$.
This requires appreciable USC mixing $c_{02}(g)$, for the Stokes pulse to be large enough
given the limitations set by dephasing $T \le T_\phi$. In other words in the USC regime 
the state $\ket{2b}$ is obtained with nearly unit probability, irrespective on $g$ 
provided it is large enough, whereas 
if mixing is insufficient there is no channel for population transfer to 
the desired target state by STIRAP. Therefore detection of $n=2$ photons in the cavity after the pulse sequence is a smoking gun for USC. Since  the best efficiency and robustness performances of 
for STIRAP are obtained for equal peak values of $\Omega_{p,s}(t)$, the amplitude of the pump 
field must be attenuated to a value such that  
$ \kappa_{p}^\Lambda :=\max[\mathscr{W}_p(t)]/\max[\mathscr{W}_s(t)] =c_{02}(g)/ c_{00}(g)$.


\begin{figure}[t!]
\centering
\resizebox{0.85\columnwidth}{!}{\includegraphics{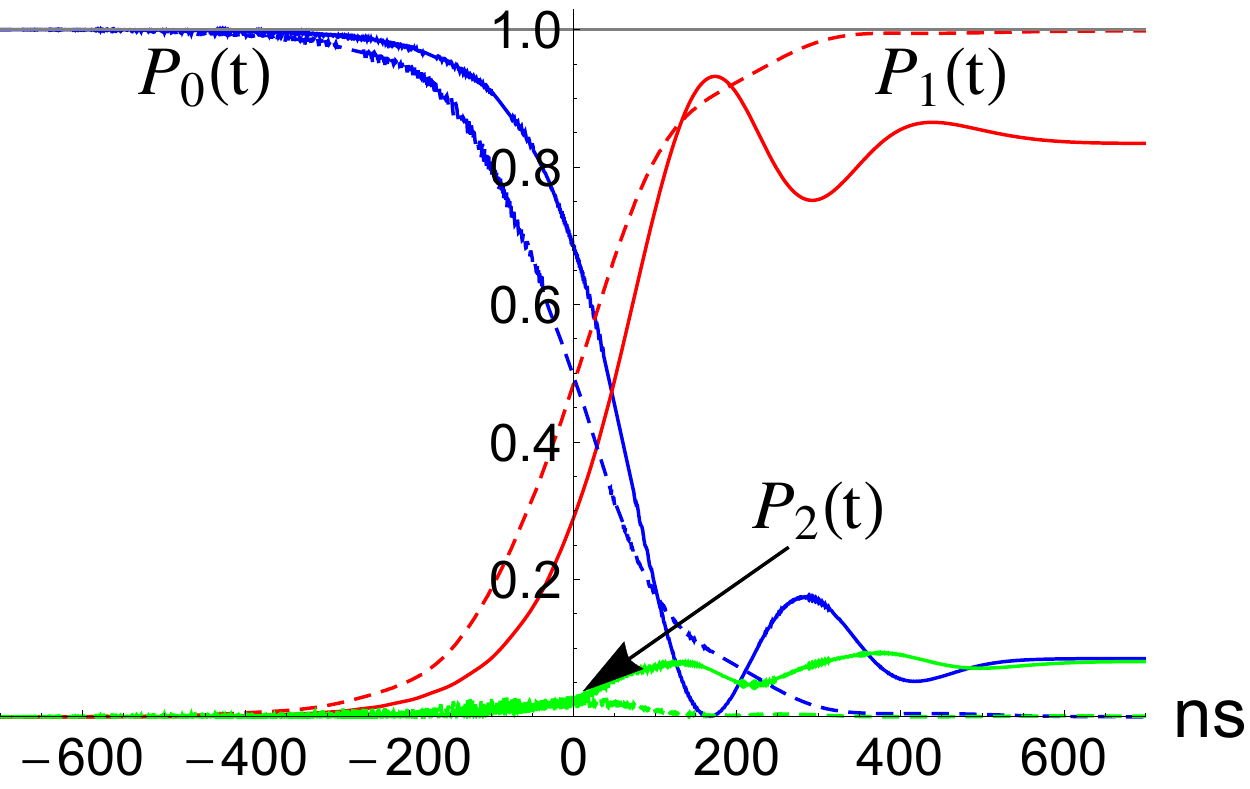}}
\caption{Coherent population transfer by STIRAP between $\ket{0} := \ket{0b}$ and $\ket{1} := \ket{2b}$, via the virtual intermediate state $\ket{2} := \ket{\Phi_0}$, as a result of USC.
Here we used a not too large value $g/\omega_c=0.2$ and figures of the external control typical of 
flux qubits~\cite{ka:210-niemczyck-natphys-ultrastrong,ka:211-bylander-natphys}. 
For this simulation 19 states were enough and considered coupling of the 
control field to all the $\ket{b}-\ket{g}$ transitions, and additional stray 
coupling also the $\ket{g}-\ket{e}$ transitions. Coherent population transfer of 
$\sim 80 \%$ is obtained (solid lines), due to partially autocompensated dynamical Stark shifts. 
Complete population transfer (dashed lines) can be achieved by an extra phase modulation, 
or by a suitable extra tone in $W_s(t)$.
\label{fig:usc2}}
\end{figure}
In order to apply this idealized procedure to realistic physical systems still several problems 
have to be addressed. We leave to 
\S\ref{sec:conclusions} a discussion of issues about the physical implementation and 
we focus on the main mathematical limitation of our model, namely that 
insufficient coupling may require a control field $W(t)$ with large Stokes amplitude 
$\max[\mathscr{W}_s(t)]$, which may address unwanted levels both in and outside the three-level 
subspace $\mathrm{span}\{\ket{0b},\ket{\Phi_0},\ket{2b}\}$. To this end we have simulated the 
idealized protocol using the full Hamiltonian $H_0+H_c(t)$ with up to 50 levels of the composite
atom-cavity system for the larger $g$ values. Stray extra terms in the coupling may induce dynamical Stark shifts of non-resonant transitions: for instance the large 
$\mathscr{W}_s(t)$ shifts the off-resonant pump transition $E_0 - E_{0b} = E_0 + \varepsilon_b 
\to E_0  + \varepsilon_b + 2 S_0(t)$, where $S_0(t) = |c_{00} \mathscr{W}_s(t)|^2/[4(E_0  + \varepsilon_b-\omega_s)]$. This is potentially detrimental for STIRAP since as a consequence 
also the transition frequency $E_{2b} - E_{0b} = 2 \omega_c \to 2 \omega_c + S_0(t)$ inducing
a stray two-photon detuning which may be the order of the effective maximum Rabi 
frequency $\max[\Omega_s(t)]$, and would completely spoil STIRAP. 
This problem can be overcome in several ways, for instance by introducing phase modulated 
pulses cancelling stray detunings analogous to what is described in \S\ref{sec:STIRAP21} 
and Ref.\cite{ka:216-distefano-pra-twoplusone}.
Actually the multilevel nature $M>3$ of the problem brings an unexpected simplification, 
since when the three-level truncation is relaxed, the Stark shifts from higher levels partly 
compensates $S_0(t)$, allowing substantial population transfer with no further correction. 
Complete population transfer can be achieved by a phase modulated control canceling the 
stray two-photon detuning induced by the dynamical Stark shifts, which depends on very few other 
levels in the spectrum. Alternatively one may use a second tone in the Stokes pulse, well detuned from 
the Stokes transition, which produces Stark shifts of reverse sign. Details of this latter 
control procedure are outside the scope of this work and will be presented elsewhere. Here
we show the results, which coincide with those of the phase modulation control, and show that
$100 \%$ efficiency is achieved (see Fig.\ref{fig:usc2}). 

In the same figure we also included 
effects of further stray control terms $H_c^\prime(t)= W^\prime(t) \,(\ketbra{g}{e}+\ketbra{e}{g})$, 
which are expected to produce stray single-photon detunings and to slightly detune 
the cavity from the natural atomic frequency, in turn decreasing the efficiency. However again 
complete population transfer is recovered by phase modulation control or by using the 
two-tone  Stokes pulse. Stray control terms corresponding to off-resonant couplings of  
states $b-e$ could be treated along the same lines, but either they are far detuned and 
thus irrelevant, or even absent due to selection rules, as in last-generation superconducting AAs. Therefore they are neglected in our analysis.

\section{Conclusions}
\label{sec:conclusions}
STIRAP is the basis of many protocols from preparation 
of superpositions~\cite{kr:201-vitanov-annurev}
to transfer of wavepackets~\cite{kr:207-kral-rmp-controladpass},
with still unexplored potentialities for quantum information 
and quantum control. Protocols with always-on fields and their generalization to 
circuit-QED architectures yield a key ingredient for several tasks 
involving individual microwave photons, from Fock state generation~\cite{ka:209-siebrafalci-prb} 
illustrated in this work, to single-photon and generation~cite{ka:213-muckerempe-pra-stirapsinglephgen}, photon conversion and their interaction. 
The key point is that as conventional
$\Lambda$-STIRAP triggers fluorescence by the absorption of an $\epsilon_2$ photon and the emission 
of an $\epsilon_2-\epsilon_1$ one, when classical fields are substituted by quantized mode of the electromagnetic field in a cavity, the generalized protocol implements photon absorption-emission 
cycles, which are a building block for processing in circuit-QED networks. 

Besides the efficiency (or the fidelity) the virtue of STIRAP-like adiabatic protocols is 
robustness against variation of the parameters, a property which makes it advantageous over Rabi cycling when larger quantum networks are considered. We also stress that phase modulated  
protocols, designed to overcome selection rules in last-generation low-decoherence AAs, leverage
on the larger stability of microwave external controls with respect to their optical counterparts. 
More elaborated versions of phase modulation allow to use superconducting AAs with the transmon
design~\cite{ka:207-koch-pra-transmon}, which nowadays exhibit the largest decoherence times, of fractions of millisecond~\cite{ka:212-rigettisteffen-prb-trasmonshapphire}.

An interesting perspective of AAs-cavity architectures is the possibility to exploit of the USC 
regime of the interaction. In our proposal STIRAP is used implement the dynamical detection of 
USC, by coherently amplifying the signal of a new channel of population transfer opened by 
USC. Guaranteeing $\sim 100 \%$ efficiency, STIRAP is advantageous with respect than using 
SEP~\cite{ka:213-stassisavasta-prl-USCSEP} since in this latter case the rate of 
two-photon production being proportional to the very small $|c_{02}(g)|^2$. 
With respect to Raman oscillations, proposed in Ref.~\cite{ka:214-huanglaw-pra-uscraman}, 
which requires careful adjusting of pulse shape and timing
also depending on the exact value of $g$, STIRAP has the advantage of robustness, i.e.
it does not depend on adjusting too many parameters to ensure faithful amplification of the 
$\ket{0b} \to \ket{2b}$ channel. 

We mention that any realistic proposal of detection 
with a three-level atom scheme, besides requiring a the coupling $g$ to the cavity 
that cannot be too weak, at the same time requires that it does not 
involve the level $\ket{b}$. Moreover a reliable mesurement scheme for Fock states in the 
cavity is also needed. These requirement are extremely severe for 
ipresent state of the art experimental systems: however STIRAP provides a unique way to overcome 
these problems, using a different scheme which will be the subject of 
a forthcoming work.

We thank C. Ciuti, R. Fazio, S. Guerin, M. Hartmann, J. Koch, Yu. Pashkin, 
M. Paternostro, S. Savasta, J.S. Tsai and A. Wallraff for useful discussions.


\end{document}